\documentclass{elsarticle}

\usepackage{latexsym}
\usepackage{float}
\usepackage[latin2]{inputenc}
\usepackage{graphicx}
\usepackage{ae}
\usepackage{aecompl}
\usepackage{setspace}
\usepackage{amsmath}
\usepackage{amssymb}
\usepackage{amsthm}
\usepackage[hidelinks]{hyperref}

\makeatletter
\def\ps@pprintTitle{%
	\let\@oddhead\@empty
	\let\@evenhead\@empty
	\def\@oddfoot{\centerline{\thepage}}%
	\let\@evenfoot\@oddfoot}
\makeatother

\begin{document}

\title{MetaHMM: A Webserver for Identifying Novel Genes with Specified Functions in Metagenomic Samples}
	
\author[p]{Balázs Szalkai\corref{cor1}}
\ead{szalkai@pitgroup.org}
\author[p,u]{Vince Grolmusz\corref{cor1}}
\ead{grolmusz@pitgroup.org}
\cortext[cor1]{Corresponding authors}
\address[p]{PIT Bioinformatics Group, Eötvös University, H-1117 Budapest, Hungary}
\address[u]{Uratim Ltd., H-1118 Budapest, Hungary}

\date{}

	\begin{abstract}
		 The fast and affordable sequencing of large clinical and environmental metagenomic datasets opens up new horizons in medical and biotechnological applications. It is believed that today we have described only about 1\% of the microorganisms on the Earth, therefore, metagenomic analysis mostly deals with unknown species in the samples. Microbial communities in extreme environments may contain genes with high biotechnological potential, and clinical metagenomes, related to diseases, may uncover still unknown pathogens and pathological mechanisms in known diseases. While the species-level identification and description of the taxa in the samples does not seem to be possible today, we can search for novel genes with known functions in these samples, using numerous techniques, including artificial intelligence tools, like the hidden Markov models (HMMs). Here we describe a simple-to-use webserver, the MetaHMM, which is capable of homology-based automatic model-building for the genes to be searched for, and it also finds the closest matches in the metagenome. The webserver uses already highly successful building blocks: it performs multiple alignment by applying Clustal Omega, builds a hidden Markov model with HMMER components of hmmbuild and uses hmmsearch for finding similar sequences to the specified model in the metagenomes. The webserver is publicly available at \url{https://metahmm.pitgroup.org}.		
	\end{abstract}

	\maketitle
	
\section{Introduction} 

By recognizing the possibilities of the advanced metagenomic analysis, more and more efforts are underway for the exploitation of the diagnostic, therapeutic and biotechnological potential of the biological information in diverse metagenomic communities. Some methods target the description of the phylogenetic composition of these metagenomes \cite{Huson2007,Huson2012, Glass2010, Wilke2013,Kerepesi2014, Kerepesi2014a}, others attempt to describe the functional assignments of the genes, encoded in the metagenomes \cite{Wilke2013,Venter2004,Krause2006,Yooseph2007,Yooseph2008,Nyiri2017}.

The bulk lists of tens of thousands of partially annotated genes, yielded by these algorithms could be used for statistical analysis of the microbiomes, but they are difficult to analyze when we are looking for genes of specific enzymatic functions in a large metagenomic data set. Here we present a new webserver, called MetaHMM, which can be applied for finding novel genes in large metagenomes of distinct origins. The webserver unites three famous and well-established bioinformatics tools in a simple, intuitive and easy-to-use automated workflow. The webserver is available at the address \url{https://metahmm.pitgroup.org}.

For the input, the user needs to specify the novel gene and the metagenome, where the novel gene is searched for. The specification of the novel gene is done by listing several, possibly homologous, known proteins with their UniProt accession numbers. The MetaHMM webserver then computes a multiple alignment from the input protein sequences, then computes an HMM profile, then searches for sequences in the metagenome, which are most similar to the HMM profile. 

For general introductions for HMMs in biology we refer to \cite{Yoon2009} and \cite{Durbin1998}. Very roughly, an HMM takes several, aligned residue sequences as input or training set, and the HMM is built for randomly generating residue sequences that are very similar to those in the training set. Next, by any residue sequence $x$, one can compute the probability that the ``well-trained'' HMM generates the sequence $x$. If the probability is high, then $x$ is accepted, meaning that it is similar enough to the training set of sequences, and if the probability is low, then $x$ is rejected, since it is not similar enough to the training residue sequences. 

The HMM uses a finite number of ``hidden'' states, in the random sequence generation: when it is in a state $A$, it may output a residue and enter some other state $B$, chosen randomly (but, usually, not in uniform distribution). This way the HMMs can (i) model conservative subsequences in the training set, and (ii) due to the random transitions, they may  recognize still unknown genes, due to the variability of the random generation/evaluation, and (iii) unlike BLAST and its clones, they can take into account the environment of the residues in the sequences aligned (they are ``context-sensitive'').

\section{Materials and Methods}

For the input, the user needs to specify several protein sequences that are similar to the sought after genes, and also the metagenome, from which the most similar sequences need to be chosen.

The workflow of MetaHMM comprises three parts. The first two parts build the hidden Markov model, the third searches for the best hits relative to the model, in the target metagenome.  The first part applies the Clustal Omega software \cite{McWilliam2013,Sievers2011a,Li2015} for performing multiple alignment, with the default Clustal parameters. The aligned sequences are given in the output as a STOCKHOLM file (.sto) for the information of the user. 

The second part constructs the HMM model; this model will be sought after in the metagenome in the third part of the workflow. For the construction we apply the {\tt hmmbuild} tool of the HMMER3 suite \cite{Eddy2011, Johnson2010}.

In the third part, using the HMM model, the short reads or the assembled sequences in the metagenome are searched for the highest scored sequences in the HMM model by the {\tt hmmsearch} program of the HMMER3 suite \cite{Eddy2011, Johnson2010}. The detailed documentation of the {\tt hmmbuild} and the {\tt hmmsearch} components can be found at \url{http://hmmer.org/documentation.html}. 

The MetaHMM server returns the best hit sequences from the metagenome, which, typically, contain short read sequences. For the identification of the whole sequences of the genes, one would need to apply a metagenomic assembly program (see, e.g., \cite{Vollmers2017}  for a critical and detailed review of the recent metagenomic assemblers).

\section{Discussion and Results}

The MetaHMM webserver can be found at the address \url{https://metahmm.pitgroup.org}. At every input field a help sign is available for a succinct information. Metagenomes can be chosen from a pre-defined list, or, alternatively, they can be uploaded to a private ftp or webserver, and the link to the uncompressed or compressed FASTA file needs to be specified. There is an upper bound of 1 GB installed for the size of this FASTA file.

The output of the MetaHMM webserver consists of the following files:
\medskip

\begin{obeylines}
\tt
input\_unaligned.fasta
The original, unaligned protein sequences comprising the model, as a FASTA file.
input\_aligned.sto
The aligned sequences of the model, as a STOCKHOLM file.
input\_profile.hmm
The HMM profile, built on the aligned sequences of the model.
output\_all.csv
All the matching domains found in the metagenomes, combined into one file.
Matches are listed by full sequence E-value ascending (best first).
output\_unaligned.fasta
All the matching domains found in the metagenomes, combined into one file.
Only those with E-value < 1e-9 are included.
output\_***.txt
Detailed search results, one file for each metagenome involved in the search.
The files are in the native output format of hmmsearch.
Those sequences which matched the model are listed after each other, ordered by E-value ascending (best first).
\end{obeylines}

\medskip

The MetaHMM webserver unites three well-known and well-established bioinformatical tools. The comparison of their power against other tools and the evaluation of those findings can be found in numerous articles, e.g., 
\cite{Eddy2011, Johnson2010,McWilliam2013,Sievers2011a,Li2015}.

\section*{Funding}
VG was partially supported by the 2017-1.3.1-VKE-2017-00013 program of National Research, Development and Innovation Office of Hungary.

\bigskip 
\noindent Conflict of Interest: The authors declare no conflicts of interest.



\end{document}